# An Empirical Study on Detecting COVID-19 in Chest X-ray Images Using Deep Learning Based Methods

Ramtin Babaeipour
RIV Lab., Department of Computer Engineering
Bu Ali Sina University
Hamedan, Iran
ramtin.babaeipour1076@gmail.com

Elham Azizi
RIV Lab., Department of Computer Engineering
Bu Ali Sina University
Hamedan, Iran
elham.azizi749@gmail.com

Hassan Khotanlou
RIV Lab., Department of Computer Engineering
Bu Ali Sina University
Hamedan, Iran
khotanlou@basu.ac.ir

*Abstract*—Spreading of COVID-19 virus has increased the efforts to provide testing kits. Not only the preparation of these kits had been hard, rare, and expensive but also using them is another issue. Results have shown that these kits take some crucial time to recognize the virus, in addition to the fact that they encounter with 30% loss. In this paper, we have studied the usage of x-ray pictures which are ubiquitous, for the classification of COVID-19 chest X-ray images, by the existing convolutional neural networks (CNNs). We intend to train chest x-rays of infected and not infected ones with different CNNs architectures including VGG19, Densnet-121, and Xception. Training these architectures resulted in different accuracies which were much faster and more precise than usual ways of testing.

## I. Introduction

Nowadays COVID19 virus has become a universal issue. The outbreak of this virus become wider over time. According to the World Health Organization, this pandemic increases exponentially every day [1]. It has caused serious damages to the global society and has led to the cancellation or suspension of many events [2]. World Health organization has also propounded that an average number of individuals that a coronavirus patient can infect is two, so early diagnosis of infected patients is a critical matter [3]. The medical community has employed many tools to satisfy the problem and has produced different test kits [4]. These test kits have shown to have 30% loss and therefore they are not accurate at all [5]. Besides, not only these testing are expensive, hard to find, and time-consuming, which are important and critical factors to avoid its spread, but also this virus has shown mutations in some cases [6]. The mutations cause different behavior of the virus and subsequently, requires different methods for detection [7]. As a result, it is important to provide a system that can solve all these issues.

CT scans and X-ray machines which have been employed among all hospitals for a long time, are widely used in many medical applications including disease detection, surgeries, etc. These machines have proved to be very practical and helpful. Chinese doctors have used chest X-rays to detect COVID-19 and achieved good results [8]. Not only these pictures state the infection but also show the tensity. Even though using X-rays has helped the diagnosis of COVID-19, this process might still encounter with human's errors and lack of necessary speed to detect this disease. In This study, we tend to process chest X-rays by deep learning and benefit good results considering that, computers can reduce human errors, as well as speeding up this process. Deep learning and computer vision have shown to be efficient in most applications of Artificial Intelligence(AI). Deep learning uses multiple layers to progressively extract higher-level features from the raw input [9]. To deliberate this, we have gathered chest X-rays of 311 people diagnosed to have COVID-19 and 311 pictures of not infected ones [10]. These X-rays then applied to our pre-trained models for classification of COVID-19 and normal cases, which provide state-of-the-art CNNs. These models have been compared to achieve best practice.

## II. Related Works

The traditional model of clinical practice consists of three steps, including diagnosis, prognosis, and treatment [11]. Deep learning has shown remarkable results in disease detection and diagnosis with the aid of large medical datasets. A study on chest X-rays in order to detect pneumonia has reached radiologist-level detection [12]. skin cancer detection with the aid of neural networks has also exhibited outstanding results [13]. There have been several studies about detecting the coronavirus by CT scans. some of these researches involve the usage of X-ray pictures and their efficiency for COVID-19 diagnosis. These studies discuss how much X-ray pictures are acceptable by the medical society. On the other hand, some researches investigate the detection and improvement of the COVID-19 virus in the context of AI. In this viewpoint most of the studies used deep learning which has concluded favorable results. These tasks are pertinent to our study.

Some researches in which X-rays and CT-scans are employed for finding COVID-19 has been done [14] [15] [16]. In these studies, they tend to identify afflicted person not by test kits, but by challenging X-rays thanks to Deep learning. Another detailed research has trained Inception Recurrent Residual Neural Network (IRRCNN) [17] model with a pneumonia dataset and applied Transfer Learning (TL) for training the same model with a COVID-19 dataset [18]. Our proposed approach tends to contrast different networks in order to find a better one.

## III. NETWORKS AND IMPLEMENTATIONS

In this paper we study three different CNNs in order to compare to find the best one that fits our data. We have used VGG-19 [19], DensNet-121 [20] and Xception [21]. All these networks are pre-trained on images from the ImageNet dataset [22].

In these networks a batch size of 8 is used. We have applied Adam optimizer with an initial learning rate of 0.001 [23]. The number of epochs has chosen to be 10. For loss computation, Binary Cross-Entropy Loss is applied [24].

*Network*

1) VGG-19

VGG-19 was proposed by K. Simonyan and A. Zisserman in the paper "Very Deep Convolutional Networks for Large-Scale Image Recognition" [19]. This model was submitted to the Image Net Large Scale Visual Recognition Challenge(ILSVRC) in 2014 and ended up getting second place in image classification task and first place in the localization task. One of the major changes in this network compared to previous ones is a smaller kernel size, where instead of 5*5 in LeNet-5 [25] and 11*11 in AlexNet [26] kernel size is reduced to the size of 3*3 which leads to increase the performance with a significant decrease in learnable parameters and lower computational complexity. VGG-19 network is 19 layers deep, consisting of 16 convolution layers for feature extraction, 3 fully connected layers, 5 MaxPool layers [27] followed by SoftMax activation for classification purpose. The downside of VGG19 is that the network architecture weights are quite large, as a result, it is *slow* to train. In this study, we use the VGG-19 architecture from the original paper, and to use a state-of-the-art image classification model we used updated weights of a pre-trained version of this network.

2) DensNet-121

In deep neural networks as back-propagation takes place, partial directives will get small and vanishing gradient problem arises [28]. Gao Huang suggests a solution to this problem with a new neural network named Densely Connected Convolutional Networks [20]. In these networks, all layers are connected directly to each other, Therefore by making input of the next layer the concatenation of all the previous layers inputs, vanishing gradient problem is solved. Previous neural networks with L layers have L connections whereas in a DenseNet there are L(L+1)/2 connections. In this model number of parameters in each layer is limited, to improve parameters efficiency. In our experiment, we used a pre-trained model of DenseNet-121 which is proposed in the Densely Connected Convolutional Networks paper [20].

3) Xception

Due to good results of inception models [29], François Chollet introduced a new model named Xception *(*Extreme version of Inception) [21]. Xception is based on inception models and Depthwise Separable Convolutions. In usual convolution applying filters across all input channels and combinations of them are done in a single step, on the other hand, Xception does this in two phases, a Depthwise [21] Convolution, and a Pointwise Convolution. In a Depthwise Separable Convolution, first, a Depthwise Convolution is done then Pointwise Convolution takes place. However, in Xception model reversed order of operations take place. Breaking that single step into these two steps reduces the number of parameters and computation time. like other models in this paper, we used a pre-trained version of the Xception model.

## IV. DATA

The dataset of chest x-rays used in this paper come from COVID-19 image data collection from Joseph Paul Cohen and Paul Morrison and Lan Dao [10]. This dataset contains x-rays of different chest diseases like Pneumonia and COVID-19. We have extracted and used Healthy and COVID-19 x-rays in our models. This dataset consists of 311 healthy RGB x-rays and 311 COVID-19 RGB ones which are then converted into 224*224 in size to fit our models. With the usage of the image data generator, we have expanded our dataset by augmentation to gain more data. After all, the obtained dataset has been split into 80% for train and 20% for test.

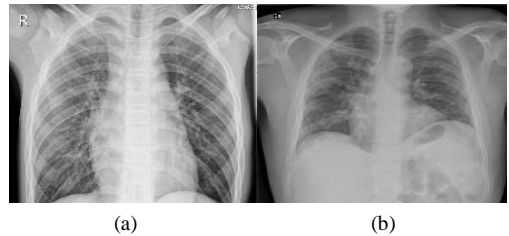

(a) (b)

Fig. 1. Images of Chest X-ray: (a) Chest X-ray of a normal case, (b) Chest X-ray of Covid-19 case

## V. RESULTS

In what follows, we conduct an experiment to compare results of training with different networks that we introduced earlier in terms of final accuracy, loss and the number of parameters. Since in our experiment the cost of having a miss-classified positive case of COVID-19 is very high we also take advantage of using three other performance measurements named precision, recall, and F1-score [30]. Accuracy is the ratio of correct predictions to all predictions. Equation (1) shows how accuracy is calculated. Next performance measurement we used in this paper is precision which demonstrates if our model predicts a positive case of COVID-19, how often it is actually a correct prediction. Equation (2) shows the precision formula. The purpose of using recall is to show of all cases that we have of positive COVID-19, how many did we label as a Positive case. Equation (3) is for calculating this measurement quantity. In the case of unbalanced samples, we use our final term named F1 Score shown in Equation (4) which is a weighted average of precision and recall. A sample result of true positive and false negative prediction, is shown in figure (2). Finally, we depicted the full training process in figure (3).

$$Accuracy = \frac{True\ Possetive + True\ Negative}{True\ Possetive + False\ Possetive + True\ Negative + False\ Negative} \quad (1)$$

$$precision = \frac{True\ Possetive}{True\ Possetive + False\ Possetive} \quad (2)$$

$$Recall = \frac{True\ Possetive}{True\ Possetive + False\ Negative} \quad (3)$$

$$F1 = 2 * \frac{Precision * Recall}{Precision + Recall} \quad (4)$$

1) *Training results for VGG-19*

Studying results of VGG-19 (Table I) shows that the VGG-19 model has reached training accuracy of 93% with training loss of 0.21. In the test part accuracy has reached 94% and test loss is 0.14. Precision in this model (Table II) shows that when our model predicted a positive case of COVID-19 about 96% of times it is actually correct. Recall displays that when it is actually a positive case of COVID-19, about 89% of times this model predicts it correctly. As a balance in these two measurements F1 scores rate is 93%. Number of trainable parameters used in this model (Table III) is less than our other models but the total number of parameters is quite large and has caused the training of VGG-19 to be very slow.

2) *Training results for Xception*

Results obtained from Training Xception (Table I) shows an improvement over VGG-19 in term of training accuracy which is 95%. Test accuracy is same as VGG-19 whereas there is a better loss in Xception. Precision factor shows same results as VGG-19 where in positive predicted cases it is 96% of times correct (Table II). There is a good improvement in the case of recall in the Xception model. Here of all COVID-19 cases about 93% of times this model predicts it correctly. F1 score is also better than the previous model. In the case of the total number of parameters, this model has the largest number of total and trainable parameters compared to our other models (Table III).

3) *Training results for DenseNet-121*

The results that emerged from training DenseNet-121 indicate that this model outperforms all the previous models in all terms. Ratio of correctly prediction to total predictions or in the other words accuracy, reaches 97% in both train and test set. Precision illustrates that 98% out of predicted positive COVID-19 cases are actually positive. Recall indicates that when it is a positive case of COVID-19 about 95% of times predict it correctly. F1 score also shows better results than our models (Table II). With 7,103,235 as the total number of parameters, DenseNet-121 has the fewest number of total parameters compared to VGG-19 and Xception hence it's training time is faster than those models (Table III).

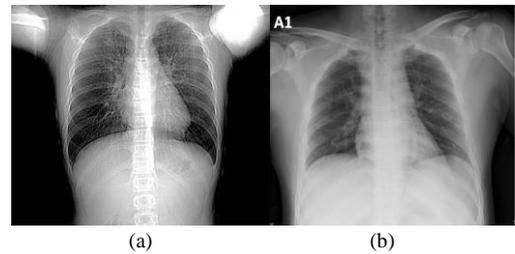

(a) (b)

Fig. 2. Chest X-ray predictions: (a) True positive, (b) False negative

TABLE I. FINAL TRAIN AND TEST SET ACCURACY AND LOSS BY DIFFERENT NETWORKS

| Networks | Accuracy And Loss Accuracy Confidence Interval: Plus or Minus 2.3% Loss Confidence Interval: Plus or Minus 0.04 | | | |
|---|---|---|---|---|
| | *train accuracy* | *test accuracy* | *train loss* | *test loss* |
| VGG-19 | 93% | 94% | 0.21 | 0.14 |
| Xception | 95% | 94% | 0.11 | 0.13 |
| DenseNet-121 | 97% | 97% | 0.08 | 0.08 |

TABLE II. NUMBER OF PARAMETERS IN DIFFERENT NETWORKS

| Networks | Performance Measures With Confidence Interval of Plus or minus 2.4% | | |
|---|---|---|---|
| | *Precision* | *Recall* | *F1 Score* |
| VGG-19 | 0.96 | 0.89 | 0.93 |
| Xception | 0.96 | 0.93 | 0.95 |
| DenseNet-121 | 0.98 | 0.95 | 0.96 |

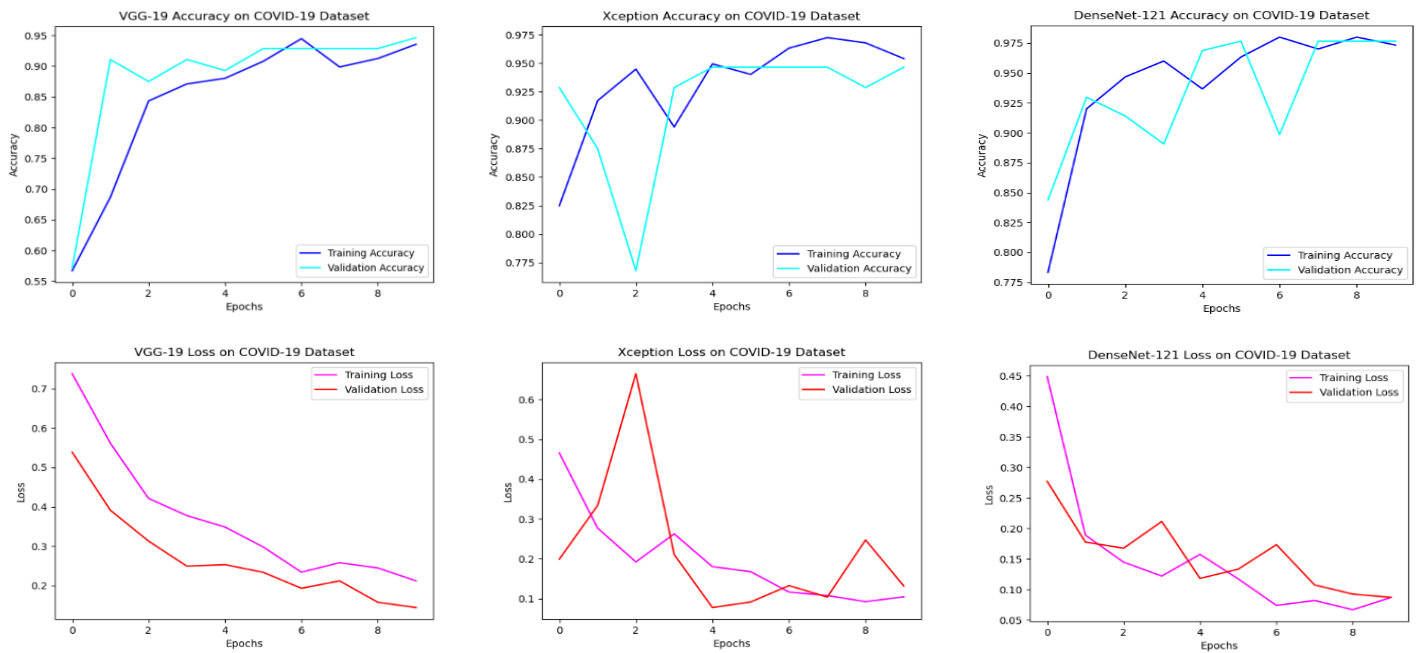

Fig. 3. Results of training for different networks. Each picture shows training process in different architectures.

TABLE III. PERFORMANCE MEASURES OF COVID-19 CASES

| Networks | Parameters | | |
|---|---|---|---|
| | Total number of parameters | Number of trainable parameters | Number of Non-Trainable parameters |
| VGG-19 | 20,057,346 | 32,962 | 20,024,384 |
| Xception | 20,992,746 | 131,266 | 20,861,480 |
| DenseNet-121 | 7,103,234 | 65,730 | 7,037,504 |

## VI. CONCLUSION

COVID-19 is a widespread infectious disease that affects millions of people worldwide right now. On account of the alarming rate of the spread of COVID-19 scientists are looking for new strategies for the diagnosis of this disease. In this paper, we have examined the performance of different CNN models to identify the best architecture for the classification of this disease. For VGG-19 results revealed that although it reached 93% accuracy, recall factor exhibits that this model mislabeled 11% of positive cases and as a result of a large number of parameters this model is slow to use. In the Xception model, we get better results with accuracy reaching 95%. Additionally, this model improves mislabeled positive cases so only 7% of times this model labels a positive case wrongly. Moreover, the observed results from experiments with DenseNet-121 indicate the best performance between these models with 97% accuracy and improvement in mislabeled positive cases where only 5% of times this model predicts it wrongly. In addition, if this model labeled a case as positive COVID-19 one, only 2% of times the label is incorrect. So with a fewer number of parameters and a faster learning pace DenseNet-121 is doing better than other models that we tested in this paper.

Despite the fact that in this study, using DenseNet-121 results in a high accuracy, we cannot conclude that using this model will result in the best outcome for classification of COVID-19 and Normal cases. We believe that with more samples, our model can be more generalized and accurate. In the future we would like to collect more samples of COVID-19 X-rays in addition to samples from other disease that can be detected from lung's X-rays like pneumonia so we can have a better classification. On top of that, we would like to collect some data about patients, like patients' ages, nationalities, and their disease backgrounds. With these new data, we would like to improve our model and make it more generalized.